\pdfoutput=1

\documentclass[final]{cvpr}

\usepackage{times}
\usepackage{epsfig}
\usepackage{graphicx}
\usepackage{amsmath}
\usepackage{amssymb}
\usepackage[version=4]{mhchem}
\usepackage{subcaption}
\usepackage[utf8x]{inputenc} 

 \usepackage{booktabs}
\newcommand{\norm}[1]{\left\lVert#1\right\rVert}
\newcommand{\innerprod}[1]{\left\langle#1\right\rangle}
\usepackage{algorithm} 
\usepackage{algorithmic} 
\usepackage{xcolor}

\usepackage[ruled,vlined,algo2e]{algorithm2e}
\usepackage{sidecap}

\usepackage[pagebackref=true,breaklinks=true,colorlinks,bookmarks=false]{hyperref}

\def\cvprPaperID{2} 
\def\confYear{CVPR 2021}

\begin{document}

\title{Generalized Unsupervised Clustering of Hyperspectral Images of Geological Targets in the Near Infrared}

\author{Angela F. Gao, Brandon Rasmussen, Peter Kulits\\\
Eva L. Scheller, Rebecca Greenberger, Bethany L. Ehlmann\\
Caltech\\
{\tt\small \{afgao, brasmuss, kulits, eschelle, rgreenbe, behlmann\}@caltech.edu}
}

\maketitle

\begin{abstract}
\vspace{-10pt}
The application of infrared hyperspectral imagery to geological problems is becoming more popular as data become more accessible and cost-effective. Clustering and classifying spectrally similar materials is often a first step in applications ranging from economic mineral exploration on Earth to planetary exploration on Mars. Semi-manual classification guided by expertly developed spectral parameters can be time consuming and biased, while supervised methods require abundant labeled data and can be difficult to generalize. Here we develop a fully unsupervised workflow for feature extraction and clustering informed by both expert spectral geologist input and quantitative metrics. Our pipeline uses a lightweight autoencoder followed by Gaussian mixture modeling to map the spectral diversity within any image.
We validate the performance of our pipeline at submillimeter-scale with expert-labelled data from the Oman ophiolite drill core and evaluate performance at meters-scale with partially classified orbital data of Jezero Crater on Mars (the landing site for the Perseverance rover).
We additionally examine the effects of various preprocessing techniques used in traditional analysis of hyperspectral imagery. This pipeline provides a fast and accurate clustering map of similar geological materials and consistently identifies and separates major mineral classes in both laboratory imagery and remote sensing imagery. We refer to our pipeline as ``Generalized Pipeline for Spectroscopic Unsupervised clustering of Minerals (GyPSUM).''
\end{abstract}
\vspace{-20pt}
\section{Introduction}
\vspace{-5pt}
Unlike a three-channel RGB image, imaging spectroscopy, or hyperspectral imaging (HSI), typically measures radiance values for hundreds of narrow wavelength bands. Material composition and physical properties determine light absorption and scattering behavior \cite{hapke_bidirectional_1981}, allowing HSI to be used for identification of materials through their spectra. Well-designed instruments can resolve individual absorption features, and images can be used to quantify material abundance and physical properties -- tasks that are generally difficult with multispectral imaging, which measures radiance for a few, typically broad wavelength bands \cite{clark_high_1990, meerdink_ecostress_2019,kokaly_usgs_2017}. Geological HSI is used for natural-hazard risk assessment and mitigation, mineral and oil exploration and production, and Earth system modeling, among other applications \cite{bioucas-dias_hyperspectral_2013}.  With increased governmental, industrial, and academic interest in and access to this technology, advanced analysis techniques for the rapidly growing wealth of hyperspectral images are becoming more valuable \cite{bioucas-dias_hyperspectral_2013}.

Typical tasks for HSI analysis of geological targets include classification, segmentation, anomaly detection, and unmixing \cite{bioucas-dias_hyperspectral_2013} and use instruments that target the visible to mid-infrared wavelengths (VIS-MIR, $\sim$400-20000 nm) for mineral, ice, and atmospheric gas identification \cite{clark_high_1990}. Semi-manual investigation is still common in these tasks \cite{bishop_surface_2018, calvin_band_2018, greenberger_compositional_2020}. One common approach for classification by expert spectral geologists is to apply knowledge of likely geologic processes occurring in the study target to isolate important known absorptions and use simple algebraic operations (``spectral parameters'') to map relative abundances of materials. Then, a system of thresholds and rules is used to classify pixels at a granularity determined by the goals of the study. Spectral parameters are also commonly used to guide basic template matching approaches of classification, such as spectral angle mapping or spectral feature fitting, which rely on extracted type spectra from the image or library spectra \cite{clark_imaging_2003}. While many attempts to partially automate this interactive and often labor-intensive workflow have been made, the same three issues are typically left unresolved: (1) analysis is time-consuming; (2) human bias can lead to false positives or negatives; and (3) expert knowledge is needed to understand the interactions between non-unique absorptions across a large space of possible materials \cite{clark_imaging_2003, calvin_band_2018}. Additionally, many geologists do not have the programming or machine learning experience required to develop more sophisticated approaches.

Supervised learning partially resolves the issues with semi-manual analysis by automating feature extraction \cite{gewali_machine_2019}. However, creation of ground-truth classification maps is expensive and requires expert input, which depends on interpretation and can be biased. Due to these challenges, only a limited collection of high-quality, fully-labeled training images for geological targets are publicly available. Large spectral libraries of minerals have been developed \cite{kokaly_usgs_2017, meerdink_ecostress_2019}, but the full range of natural spectra is large \cite{thompson_large_2017} and libraries do not contain the spatial context that exists in HSI. There are infinitely many naturally-occurring infrared  spectra because they are combinations of pure mineral spectra which are modified further by varying abundances and physical properties. Thus, large spectral variability and limited training data availability make generalizable supervised classification a difficult task in geological HSI analysis.

In this work, we develop a novel autoencoder-based (AE) feature extraction technique coupled with Gaussian mixture modeling (GMM) for fully unsupervised classification of single HSI images in the near-infrared (NIR, $\sim$ 1000-2500 nm) . In addition, we quantify the effects of traditional preprocessing methods on the pipeline. We employ both quantitative metrics and qualitative expert interpretation to determine algorithmic performance on two geological HSI datasets, which include a labelled laboratory image from the Oman Drilling Project \cite{kelemen_2020} and satellite images from the Compact Reconnaissance Imaging Spectrometer for Mars (CRISM) \cite{murchie_compact_2007}, including Jezero Crater, where the Perseverance rover recently landed (Section \ref{background}).

\vspace{-12pt}
\section{Previous Works} \label{prev_work}
\vspace{-5pt}
Unsupervised clustering of HSI is challenging for three reasons: (1) Noise profiles vary widely depending on target, imaging conditions, instrumentation, and calibration, and can have distinct spatial or spectral structure \cite{ ceamanos2010spectral, leask2018challenges, murchie_compact_2007}; (2) Performing dimensionality reduction of HSI data without losing important information is difficult \cite{thompson_large_2017}; (3) Many clustering approaches are computationally intensive and rely on distance metrics which become less meaningful as dimensionality increases \cite{murphy_unsupervised_2018, yadav_2019}. 

A commonly used unsupervised clustering algorithm is principal component analysis (PCA) and k-means clustering \cite{rodarmel2002principal}. However, k-means favors equal sized clusters (typically a poor assumption for geological targets) and PCA works well in the case of well-separated, convex clusters in the original space, which is uncommon in hyperspectral data \cite{prml}.  Deep Embedded Clustering (DEC) \cite{xie_unsupervised_2016} and Sparse Manifold Clustering and Embedding (SMCE)  \cite{elhamifar_2011} fail to consistently outperform PCA + k-means in HSI datasets \cite{yadav_2019}. Spectral-Spatial Diffusion Learning (DLSS) \cite{murphy_unsupervised_2018} is a state-of-the-art unsupervised learning algorithm for HSI \cite{yadav_2019}, but does not scale well with respect to memory and is only applicable to small datasets \cite{murphy_unsupervised_2018, yadav_2019}. Mou et al. proposed a Conv–Deconv (convolutional/encoder - deconvolutional/decoder) network for unsupervised spectral-spatial feature learning, but it requires labels for fine-tuning \cite{mou_unsupervised_2018}. Nalepa et al.\ \cite{nalepa_unsupervised_2019} used a KL divergence based objective to learn parameters for clustering. However, this method does not outperform GMM-based methods on the Salinas Valley dataset, which is most similar to the Oman and CRISM datasets, and it is significantly slower than both k-means and GMM \cite{nalepa_unsupervised_2019}. Infinite mixture of infinite GMM, I$^2$GMM, is an unsupervised method that has been shown to be successful on CRISM data \cite{dundar_2016, leask_2018}. This method results in component features that are more difficult to interpret in comparison to AE extracted features \cite{leask_2018}. 

Automated determination of the number of endmembers is crucial to successful HSI analysis. The unsupervised Hyperspectral Signal identification by minimum error (HySime) algorithm for endmember optimization infers the signal subspace in hyperspectral imagery \cite{bioucas-dias_hyperspectral_2008}, which yields comparable or superior results compared with Harsanyi-Farrand-Chang (HFC) \cite{harsanyi_determining_1993} and Noise-Whitened HFC \cite{chang_estimation_2004} eigen-based Neyman–Pearson detectors. HySURE \cite{rasti_2015} outperforms HySime for low SNR synthetic data settings but is less consistent on real datasets. Thus, we use HySime to determine the optimal number of clusters to produce. 

To best address the three issues outlined above, we develop an AE-GMM methodology with optional post-processing (GMM+) with HySime optimization of endmembers. We directly compare the performance of this new methodology, which we call GyPSUM, to semi-manual workflows and the industry standard PCA and k-means methodology.

\section{Background} \label{background}

\subsection{Autoencoder} \label{AE}
Autoencoders (AE) \cite{vincent_autoencoder} are a feature-extraction technique that consists of an encoder network $E_{\phi}$ and decoder network $D_{\theta}$ parameterized by $\phi$ and $\theta$ respectively. The encoder $E_{\phi}$ maps inputs $\mathbf{x} \in \mathbb{R}^N$ to a latent representation $\mathbf{z} \in \mathbb{R}^D$. For a single hidden layer encoder, $\mathbf{z}$ is given by
\begin{align}
    \mathbf{z} = \mathbf{W_2}\sigma(\mathbf{W_{1}}\mathbf{x} + \boldsymbol{\beta}_{1}) + \boldsymbol{\beta}_2 \label{eq:encoder}
\end{align}
where $\mathbf{W}_1$ and $\boldsymbol{\beta}_1$ are the weights and the bias vector of the hidden layer, $\mathbf{W}_2$ and $\boldsymbol{\beta}_2$  are the weights and the bias vector of the output layer, and $\sigma$ represents the nonlinear activation function.
The encoded representation $\mathbf{z}$ is used to produce a reconstruction $\hat{\mathbf{x}}$ through the decoder $D_{\theta}$. For a single hidden layer decoder, $\hat{\mathbf{x}}$ is given by
\begin{align}
    \mathbf{\hat{x}} = \mathbf{V_2}\sigma(\mathbf{V_{1}}\mathbf{z} + \boldsymbol{\gamma}_{1}) + \boldsymbol{\gamma}_2 \label{eq:decoder}
\end{align}
where $\mathbf{V}_1$ and $\boldsymbol{\gamma}_1$  are the weights and bias vector of the hidden layer, $\mathbf{V}_2$ and $\boldsymbol{\gamma}_2$  are the weights and bias vector of the output layer, and $\sigma $ represents the nonlinear activation function. Typically, the loss function is a reconstruction loss $\norm{ \mathbf{x}-\mathbf{\hat{x}}}_2$.

\subsection{Expectation Maximization Clustering Models}
Expectation maximization (EM) \cite{prml} methods calculate the maximum likelihood estimate (MLE) parameters $\boldsymbol{\theta}$ (independent of the decoder $D_{\theta}$ parameters) of a statistical model. This model depends on unobserved latent variables $\mathbf{Z}$ for a dataset $\mathbf{X}$ and likelihood function $L(\boldsymbol{\theta}; \mathbf{X}, \mathbf{Z}) = p(\mathbf{X}, \mathbf{Z}|\boldsymbol{\theta})$. In particular, EM clustering algorithms solve for the optimal clustering given a number of clusters and cluster probability distribution. 

Maximizing the marginal likelihood function
$L(\boldsymbol{\theta}|\mathbf{X}) = \int_{\mathbf{Z}}p(\mathbf{X}, \mathbf{Z}|\boldsymbol{\theta}) d\mathbf{Z}$ is often intractable. Thus, EM algorithms iteratively solve the marginal likelihood. These steps $t$ are repeated until convergence:
\newline
(1) Defining 
\begin{align}
    Q(\boldsymbol{\theta}|\boldsymbol{\theta}^{(t)}) = \mathbb{E}_{\mathbf{Z}|\mathbf{X}, \boldsymbol{\theta}^{(t)}}[\log L(\boldsymbol{\theta}; \mathbf{X}, \mathbf{Z})]. \label{eq:Estep}
\end{align}
(2) Improved estimates of $\boldsymbol{\theta}$ are computed through 
\begin{align}
    \boldsymbol{\theta}^{(t+1)} = \arg\max_{\boldsymbol{\theta}^{(t)}}Q(\boldsymbol{\theta} | \boldsymbol{\theta}^{(t)}). \label{eq:Mstep}
\end{align} 
GMM \cite{prml} is an example of an EM clustering algorithm with a Gaussian distribution for the cluster probability model. Unlike GMM, k-means does not optimize a probabilistic model. This biases k-means towards equal sized clusters.

\subsection{Quantitative Metrics} \label{background_metrics}
We use multiple unsupervised cluster-separation metrics for evaluation. The Calinski-Harabasz (CH) index \cite{calinski_harabasz} for data $E$ with $n_E$ pixels and $k$ clusters is
\begin{align}
    s = \frac{\text{tr}(B_k)}{\text{tr}(W_k)} \times \frac{n_E - k}{k-1} \label{eq:CH}
\end{align}
where 
$$W_k = \sum_{q=1}^k \sum_{\mathbf{x} \in C_q} (\mathbf{x}-\mathbf{c}_q)(\mathbf{x}-\mathbf{c}_q)^T$$
$$B_k = \sum_{q=1}^k n_q(\mathbf{c}_q - \mathbf{c}_E)(\mathbf{c}_q-\mathbf{c}_E)^T$$
with $C_q$ as the set of points in cluster $q$, $\mathbf{c}_q$ the center of cluster q, $\mathbf{c}_E$ the center of data $E$, and $n_q$ the number of points in cluster $q$. CH scores are higher when clusters are dense and well-separated but penalize non-convex clusters.

The Davies-Bouldin (DB) index \cite{davies_bouldin} is defined by 
\begin{align}
    b = \frac{1}{k} \sum_{i = 1}^k \max_{j} R_{ij} \label{eq:DB}
\end{align}
where $$R_{ij}=\frac{s_i + s_j}{d_{ij}}$$ is a cluster similarity measure between clusters $i$ and $j$. $s_i$ is the cluster diameter and $d_{ij}$ is the distance between the centroids of clusters $i$ and $j$. This score is lower when clusters are dense and well-separated, unlike the CH index.

\subsection{Oman Dataset}
Ocean crust is typically technologically challenging to drill and investigate in situ. However, the Oman ophiolite was tectonically thrust on top of stable continental crust \cite{searle_1999, Glennie_1973} and it is the subject of the International Continental Scientific Drilling Program's Oman Drilling Project \cite{kelemen_2020}. Our first dataset consists of a few images selected from $>$ 31 TB of drill core images ( $>$ $3.2$ km of imaged slices of $\sim$ 5 cm diameter extracted rock core). Images were collected in-lab with Caltech's custom imaging spectrometer with a spectral resolution of $\sim$6 nm and a spatial resolution of $\sim$0.25 mm/pixel at wavelengths from 900 to 2600 nm. These high SNR images have high spectral variance compared with typical imagery, but contain rare, low spatial occurrence mixtures that are hard to isolate.

\subsection{CRISM Dataset}
CRISM is a push-broom visible/infrared imaging spectrometer aboard the Mars Reconnaissance Orbiter that has been orbiting Mars since 2006. It has collected hundreds of thousands of images of the planet's surface in varying image modes ranging from coarse multispectral imagery to high spatial resolution targeted imagery at 18-40 m/pixel at spectral sampling of $\sim$6.5 nm across the wavelength range of 362-3920 nm \cite{murchie_compact_2007}. HRL000040FF ($\sim$40 m/pixel) is an image of the Jezero Crater rim and an ancient crater lake delta that is the main target of exploration for the recently landed Perseverance rover \cite{farley_mars_2020,goudge_assessing_2015}. We also include FRT0000634B\footnote{included in the supplementary materials} ($\sim$18 m/pixel), an image from the Claritas Rise that shows evidence of hydrothermal alteration \cite{ehlmann_geologic_2010}. Challenges to clustering include systematic cross-track dependent noise, pixels that always represent only mixtures due to coarse spatial resolution, subdued absorptions indicative of minerals, and atmospheric residual absorptions which overlap with mineralogically important absorption features. 

\section{GyPSUM Method} \label{methods}
Our pipeline consists of preprocessing detailed in Algorithm \ref{alg:preprocess}, feature extraction and clustering detailed in Algorithm \ref{alg:main}, and an optional post-processing step. 


\subsection{Preprocessing}

\vspace{-5pt}
We implement two similar preprocessing workflows in the case of laboratory imaging versus orbital CRISM imagery detailed in Algorithm \ref{alg:preprocess}. We begin with reflectance data and clip values between 0 and 1 to remove nonphysical outlier spikes before trimming the data in the spectral dimension to between 1050 and 2550 nm; this spectral subset contains the absoprtion features of greatest interest for our studies. We normalize each spectrum with the per-pixel $\ell_2$ norm and reshape the data to a flat vector of pixels. We then apply a mask created from previous work to remove background material. While not necessary for generalized implementation, we find that this increases the total variance captured by a fixed dimension of latent components. Finally, we optionally divide out a linear convex hull computed for each spectrum, a technique (continuum removal, CR) commonly used to visually interpret subtle absorption features \cite{clark_reflectance_1984}.
\newcommand{\m}{\-\qquad}
\vspace{-4pt}
\begin{algorithm}[H]
\KwData{ $\mathbf{X}$ HSI cube n $\times$ m $\times $ w, optional mask $\mathbf{M}$}
\KwResult{$\hat{\mathbf{X}}$ preprocessed vectorized image p $\times$ w}
$\mathbf{X} \gets $ ClipReflectance($\mathbf{X}$) \ \ \ \% \textit{clip values from 0 to 1}\;
 
\If {CRISM}{
$\mathbf{X} \gets$ RatioImage($\mathbf{X}$) \quad \% \textit{divide by ratio spectrum}
}
$\mathbf{X} \gets $ ClipWavelengths($\mathbf{X}$) \   \% \textit{clip from 1050 to 2550 nm}\;

$\mathbf{X} \gets $ ${\mathbf{X}}./{\norm{\mathbf{X}}}_2$ \quad \quad \quad \quad \ \ \  \% \textit{per pixel normalization}\;

\If{mask}{
    $\mathbf{X} \gets $ mask($\mathbf{X}$, $\mathbf{M}$) \quad \ \ \   \% \textit{mask out unwanted pixels}
}
\If {not CRISM}{
    $\mathbf{X} \gets$ RemoveContinuum($\mathbf{X}$) \quad \% \textit{get convex hull}
}    
 $\hat{\mathbf{X}}$ = Flatten($\mathbf{X}$)
 \caption{Preprocessing}\label{alg:preprocess}
\end{algorithm}
 
In the case of orbital CRISM imagery, we start with MTR3 products from the Planetary Data System \cite{seelos_crism_2011}, considered the highest-fidelity publicly available CRISM imagery. A sophisticated set of empirical and statistical corrections have been pre-applied to this data to remove spikes, correct for imaging geometry and gimbal motion, and remove atmospheric contamination to retrieve approximate surface reflectance. It is noteworthy that the method used for atmospheric correction is imperfect and leaves considerable \ce{CO2} absorption residuals \cite{mcguire_improvement_2009}. One common way that atmospheric residuals and systematic cross-track dependent noise is removed in CRISM imagery is through spectral ratioing with bland material within the image, which also emphasizes minor mineral components \cite{ehlmann_geologic_2010,bishop_surface_2018}. We optionally manually develop a ratio spectrum from a mean of many bland pixels for each image  and divide it out of every pixel of the image. 
\vspace{-4.6pt}

\begin{SCtable*}
    \centering

\begin{tabular}{l|cc|cc|cc}
\toprule
   & \multicolumn{2}{c|}{F1} &
  \multicolumn{2}{c|}{NMI} &
  \multicolumn{2}{c}{ARI}  \\
  \hline
{} &  NCR &    CR &  NCR &    CR &   NCR &    CR  \\
\hline
11 Cl. &     \textbf{0.260} & 0.258 &     0.254 & 0.289 &     0.122 & 0.185 \\
20 Cl. &     0.243 & 0.236  &     0.282 & 0.313 &     0.109 & 0.149 \\
GMM+ &     0.157 & 0.149 &     0.332 & \textbf{0.402} &     0.170 & \textbf{0.221} \\
PCA+  &     0.132 & 0.129 &     0.223 & 0.238 &     0.148 & 0.181 \\
\bottomrule
\end{tabular}

    \caption[]{\small \textbf{Supervised Metrics: Oman.} F1, Normalized Mutual Information (NMI), and Adjusted Rand Index (ARI) using expert labels (11 classes). Clusters generated using non-CR (NCR) and CR embeddings from GMM (11 and 20 clusters), GMM with twice the HySime output (52) and post-processed to 11 (GMM+, full GyPSUM pipeline), and PCA (20 components) + k-means (52 clusters) + post-processing to 11 (PCA+). Best score for each metric is bolded.}
    \label{tab:oman_supervised_metrics}
\end{SCtable*}

\subsection{Feature Extraction}

In order to develop a general approach that performs well across different noise settings, we use an autoencoder framework, which has been shown to perform well at denoising \cite{vincent_autoencoder}. We train on a per-image basis to leverage the inherent mineralogical similarity within an image target. For each image, we train a lightweight autoencoder to learn a per-pixel embedding that is of a lower dimension $d$ than the input image space $w$. The training data for the image-specific autoencoder consists of the entire hyperspectral image of interest with $p$ pixels with $w$ channels each. The input to the autoencoder is a single pixel $\mathbf{x} \in \mathbb{R}^w$, which is a single $w$ channel spectrum. The encoder network $E_{\phi}$, parameterized by $\phi$, generates learned features $\mathbf{z} \in \mathbb{R}^d$. Then, the decoder network $D_{\theta}$, parameterized by $\theta$, generates a reconstructed spectrum $\mathbf{\hat{x}} \in \mathbb{R}^w$. The dimensionality $d$ of the learned feature space $ \mathbf{Z} \in \mathbb{R}^d$ is determined by HySime \cite{bioucas-dias_hyperspectral_2008}. The learned features $\mathbf{z}$ are then used for clustering. 
\vspace{-8pt}

\begin{algorithm}[H]
\SetAlgoLined
 \KwData{ $\hat{\mathbf{X}}$ vectorized image p $\times$ w, optional spectral angle threshold $\lambda$, optional embedding size $d$, optional number of clusters $k$} 
 \KwResult{$\mathcal{C}$ clusters p $\times$ $k$}
    \If{$d$ is not given}{$d \gets$ HySime($\hat{\mathbf{X}}$) }
     $\mathbf{Z} \gets $ Autoencoder($\hat{\mathbf{X}}$ , $d$) \% \textit{train with preprocessed data}\;
     
     \If{$k$ is not given}{$k \gets 2d$} 
     $\mathcal{C} \gets $ GaussianMixtureModel($\mathbf{Z}$, $k$)\;

     \If {PostProcess}{
    $\mathcal{C} \gets$ PostProcess($\mathcal{C}$, $\lambda$)
}
 \caption{Feature Extraction and Clustering}\label{alg:main}
\end{algorithm}

We use a lightweight two-hidden-layer encoder and decoder architecture for training efficiency as well as implicit regularization. We use rectified linear unit (ReLU) as the activation function. Using ReLU instead of output-constrained activation functions, like sigmoid and hyperbolic tangent, allows for faster convergence \cite{hayou_2019}. We use Adam \cite{zhang_dive_2020} as our optimizer with a learning rate of $10^{-3}$ and train until convergence. Instead of the typical mean squared error reconstruction loss, we use the spectral angle (SA) between the input $\mathbf{x}$ and reconstruction $\mathbf{\hat{x}}$, which is defined by \begin{align}
\text{SA}(\mathbf{x}, \mathbf{\hat{x}}) =\arccos\left( \frac{\innerprod{ \mathbf{x}, \mathbf{\hat{x}}}}{\norm{\mathbf{x}}_2\norm{\mathbf{\hat{x}}}_2}\right). \label{eq:SA}
\end{align} Using spectral angle allows for invariance to relative magnitude, which results in decreased sensitivity to network initialization, and increases the cost in small scale variation, which helps to capture small features in the spectra. Learning a per-pixel embedding for each image allows for robustness to various mineral abundance distributions and varying systematic instrumental and environmental noise.

\subsection{Clustering}
We cluster the learned embeddings from the autoencoder on a per-pixel basis using a Gaussian mixture model with full covariance matrices. Unlike k-means, GMMs are robust to small clusters and can separate clusters that are not well-separated in space. Since the features can be interpreted as corresponding to different minerals, enforcing feature independence would artificially prevent mixtures of minerals to be identified. To determine the number of clusters, we use twice the number estimated by HySime \cite{bioucas-dias_hyperspectral_2008} on the spectral data since HySime tends to underestimate the number of distinct endmembers \cite{rasti_2015}. We can then use post-processing to reduce the number of clusters to a manageable number. 



\vspace{-3pt}
\subsection{Post-Processing}
Because we fix the number of classes before clustering to twice the estimate from HySime \cite{bioucas-dias_hyperspectral_2008}, we optionally post-process output clusters by merging redundant classes based on their mean spectra. We use spectral angle as a similarity metric between mean spectra of each cluster. We iteratively combine pairs of clusters with the smallest spectral angle between cluster means until the minimum spectral angle between cluster means exceeds a user-defined threshold $\lambda$.

\subsection{Evaluation}
Classification of geological materials is interpretive by nature, with necessary class specificity dependent on the application. For example, initial expert classification maps for the Oman core data did not differentiate two spectrally distinct zeolite minerals because they did not inform the scientific goal of determining trends in hydration, formation temperatures, and water chemistry with depth. To assess performance of varying preprocessing techniques, autoencoder architectures, embedding sizes, and number of clusters, we apply visual qualitative analytical methods and compare against PCA + k-means, which is used widely in software for unsupervised HSI classification. We hand-select regions of interest (ROIs) in each image covering the range of important mineralogical diversity determined by two spectral geologists in the case of the laboratory data, and through both a partially expert-labelled image and literature results across many publications for the CRISM data \cite{viviano-beck_compositional_2017,horgan_mineral_2020,goudge_assessing_2015,ehlmann_geologic_2010,tarnas_orbital_2019}. We visually assess each class present within the ROIs to determine if the class is consistently mapping similar material across a large subset of the image. Additionally, we assess whether cluster means of the classes comprising the pixels from the ROIs contain the absorption features representative of the mineralogy, and if mixing with other distinct mineralogy is muted.

In addition to the qualitative evaluation, we employ several quantitative metrics to evaluate our methods. We compute metrics that evaluate the separation and density of clusters for each variation of our pipeline. In particular, we use unsupervised metrics (CH index (Eq.\  \ref{eq:CH}) and DB index (Eq.\ \ref{eq:DB})) and supervised metrics (F1, Normalized Mutual Information (NMI), and Adjusted Rand Index (ARI)) \cite{Rabbany_2017} for the Oman dataset, which has spatially complete expert labels available. 

\begin{figure*}[ht!]
    \centering
    \captionsetup[subfigure]{justification=centering}
    \begin{subfigure}[t]{0.21\textwidth}
        \centering
        {\includegraphics[width=\textwidth]{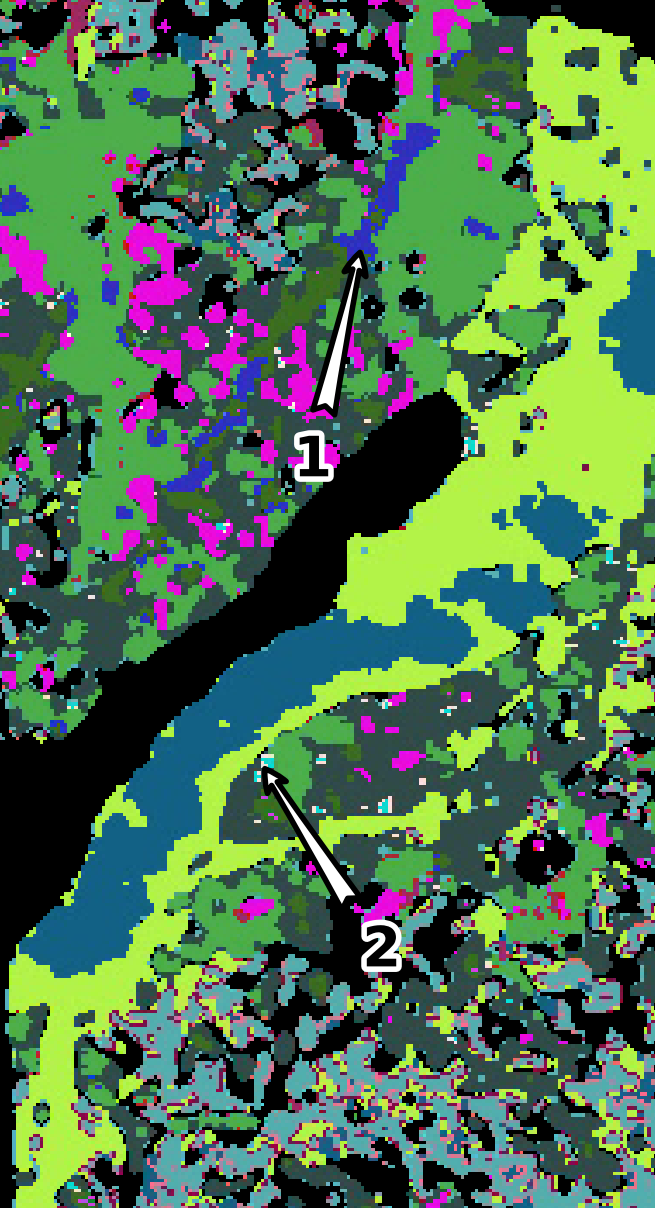}}
        \caption[]
        {{\small Expert Labels}}    
        \label{fig:oman_crop_gt}
    \end{subfigure}
    \hfill
    \begin{subfigure}[t]{0.21\textwidth}
        \centering
        {\includegraphics[width=\textwidth]{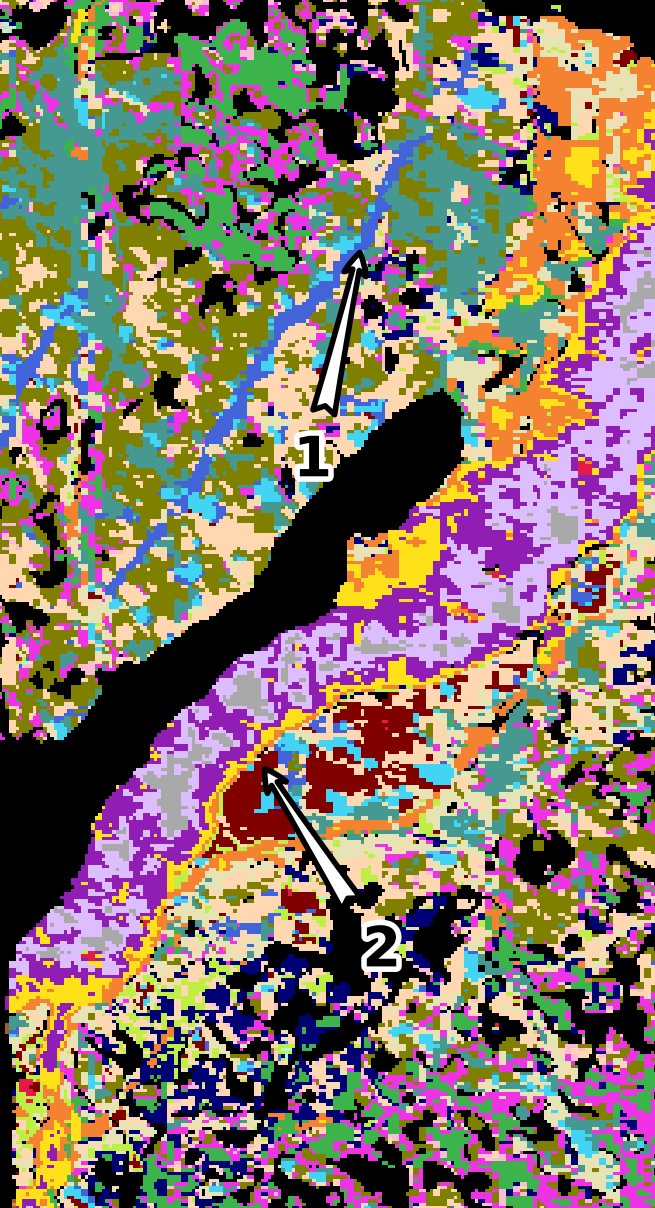}}
        \caption[]
        {{\small PCA + K-Means}}
        \label{fig:oman_crop_pca}
    \end{subfigure}
    \hfill
    \begin{subfigure}[t]{0.21\textwidth}   
        \centering 
        {\includegraphics[width=\textwidth]{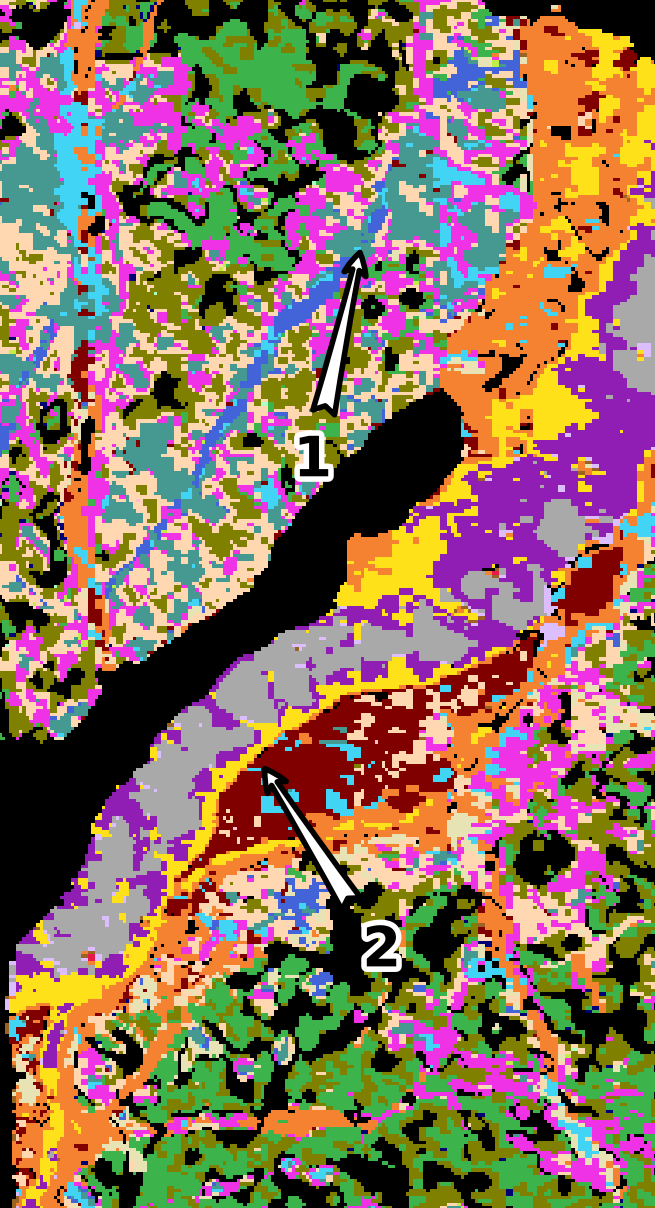}}
        \caption[]
        {{\small AE + GMM on\\ Spectral Data}}    
        \label{fig:oman_crop_ae}
    \end{subfigure}
    \hfill
    \begin{subfigure}[t]{0.21\textwidth}   
        \centering 
        {\includegraphics[width=\textwidth]{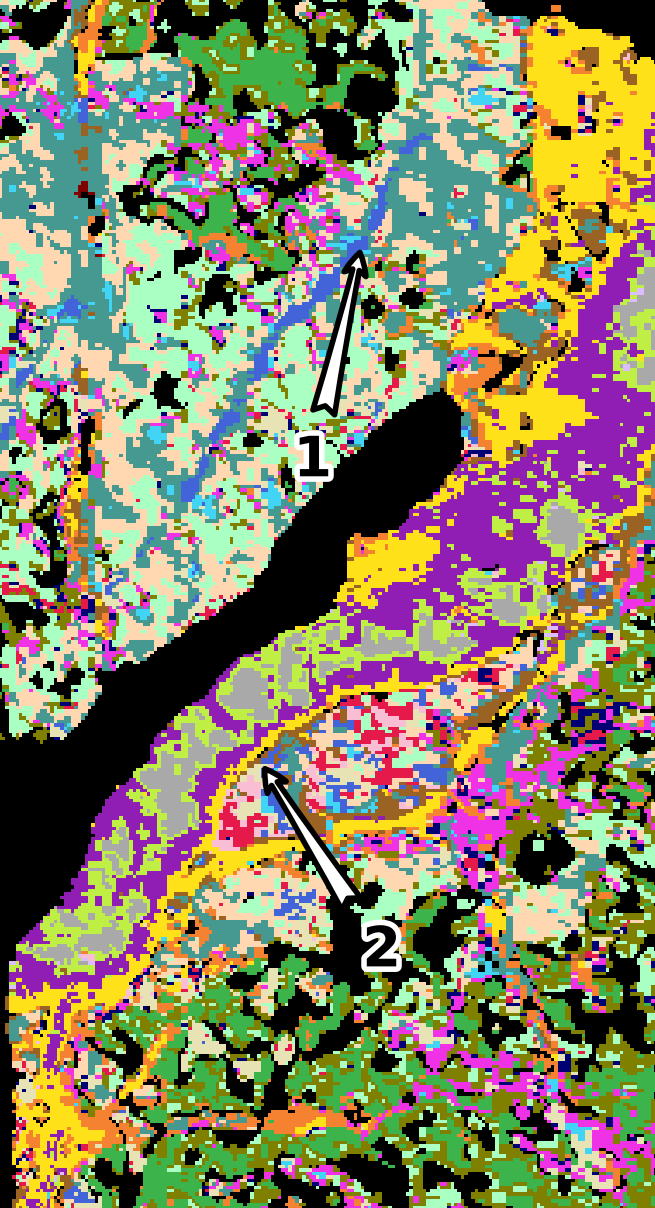}}
        \caption[]
        {{\small AE + GMM on  CR Data}}
        \label{fig:oman_crop_cr}
    \end{subfigure}
    \caption[ ]
    {\small \textbf{Oman core ROI with different methods} (c) and (d) clearly capture the epidote vein (dark green {\color[HTML]{3A6E1B}$\blacksquare$} below 1 in (a), cobalt blue {\color[HTML]{4363D8}$\blacksquare$} in (c) (d)),
    while separating distinct prehnite/epidote mixtures (dark blue {\color[HTML]{1837CD}$\blacksquare$} in (a), blue-green {\color[HTML]{469990}$\blacksquare$} in (c) (d), labeled 1). PCA + k-means (b) fails to identify this distinct mixture. All methods struggle with distinctly mapping amphibole (labeled 2, turquoise {\color[HTML]{0BDAD5}$\blacksquare$} in (b)).
    Full size core images are available in the supplement.}
    \label{fig:oman-crop}
\end{figure*}

\begin{figure}
    \centering
    \begin{subfigure}[t]{0.31\textwidth}  
        \centering 
        \includegraphics[width=0.49\textwidth]{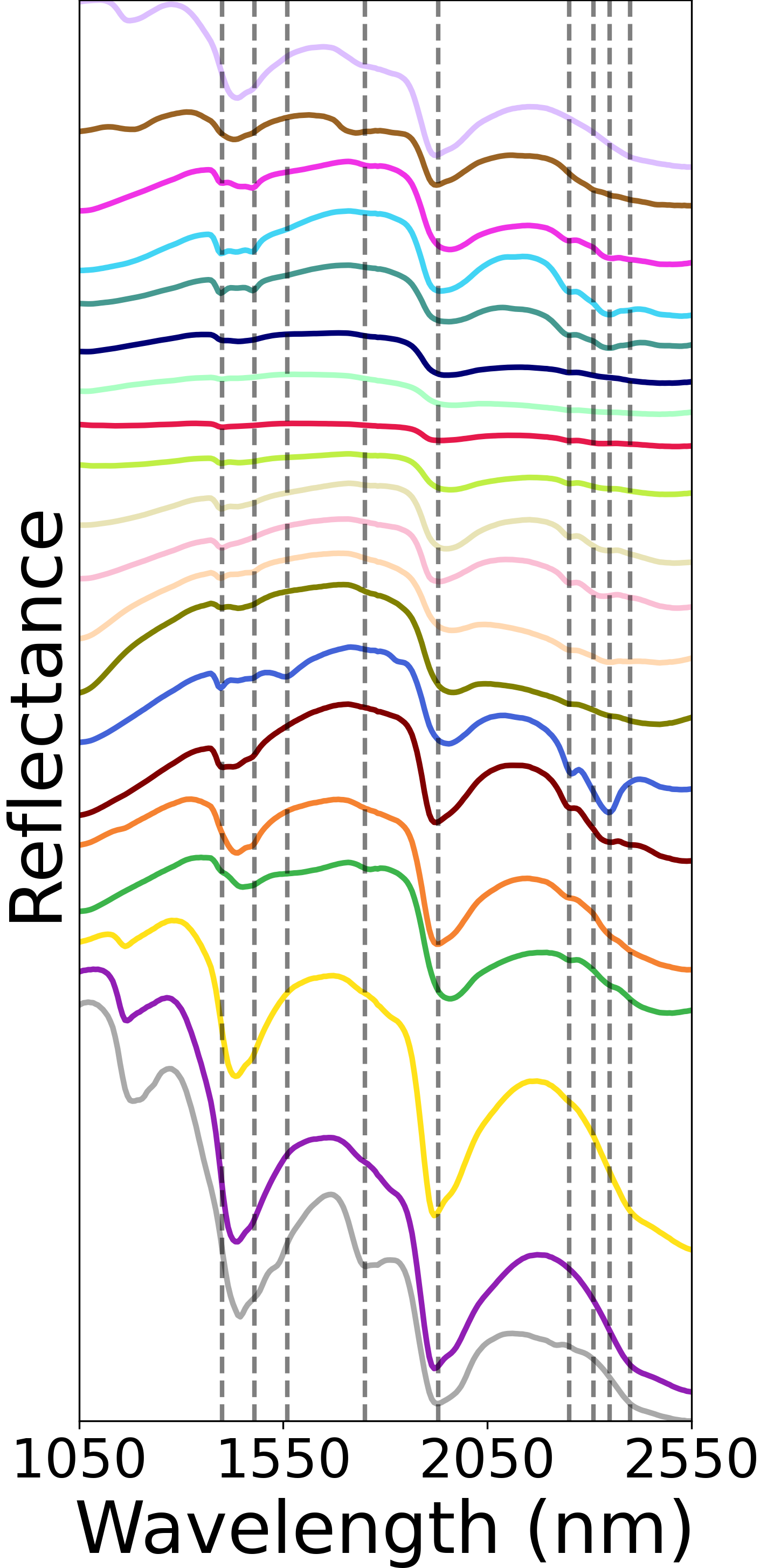}
        \includegraphics[width=0.49\textwidth]{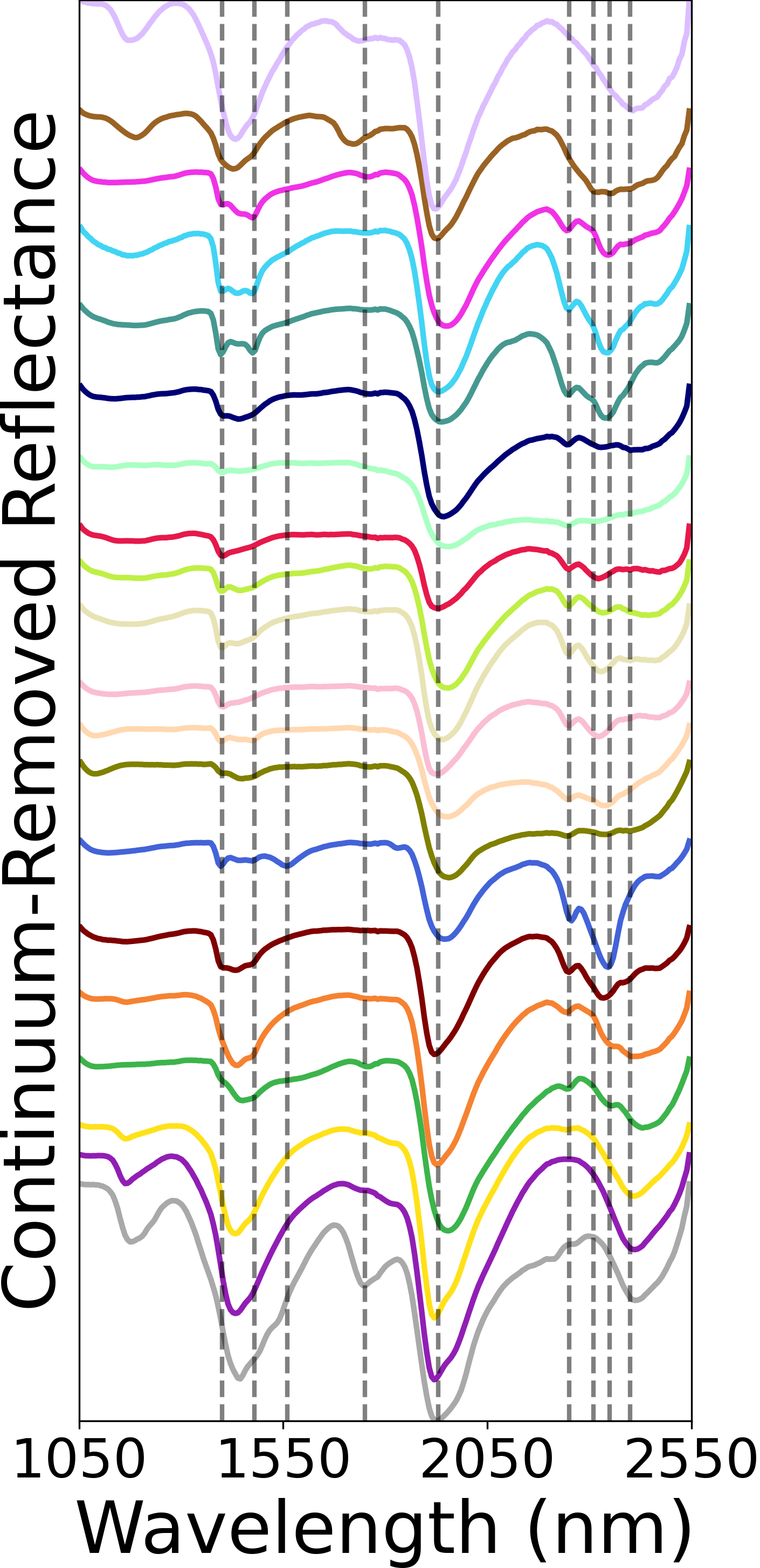}
        \caption[]{}
        \label{fig:oman_quint_b}
    \end{subfigure}
    \hfill
    \begin{subfigure}[t]{0.155\textwidth}   
        \centering 
        \includegraphics[width=0.98\textwidth]{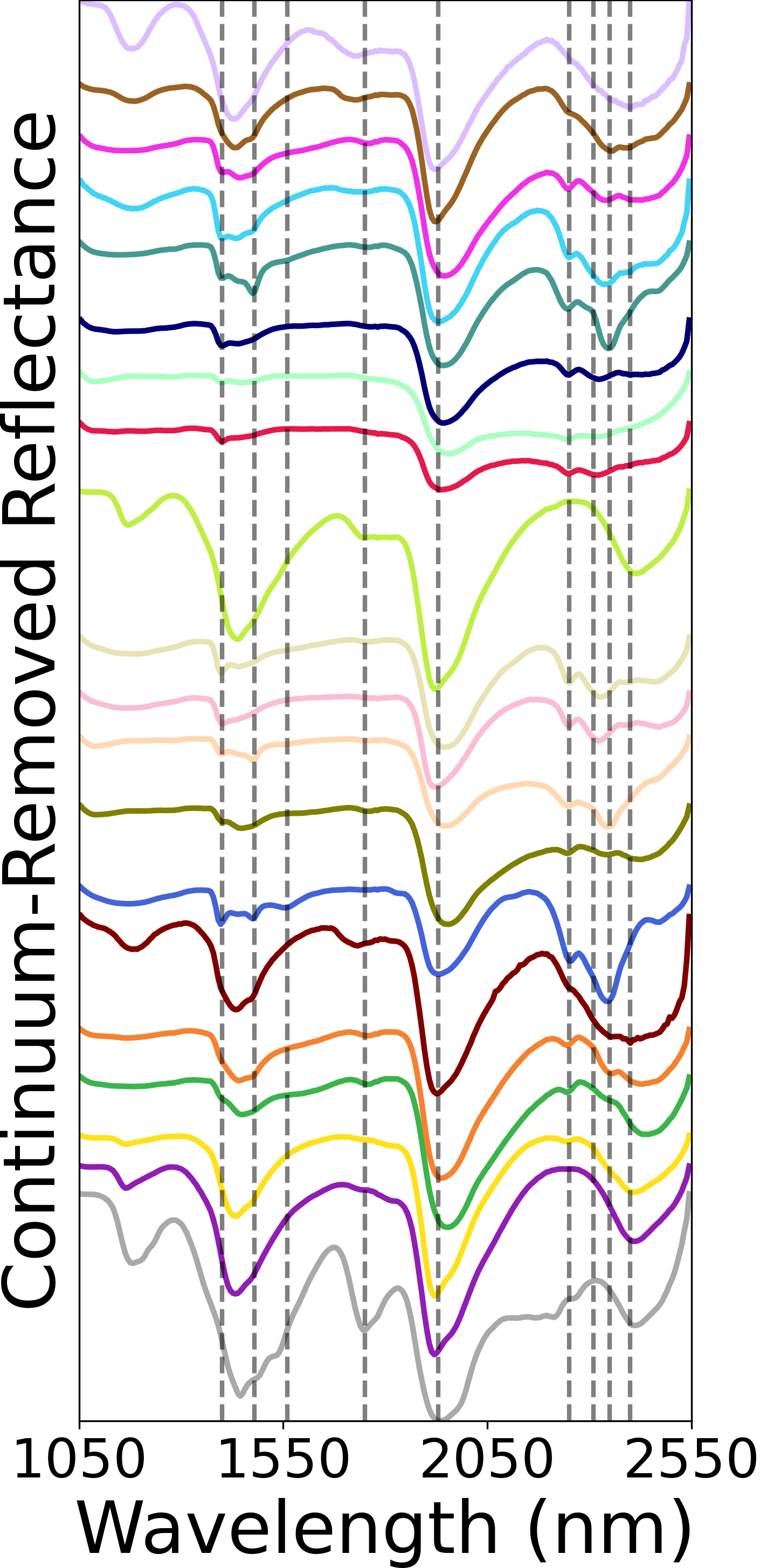}
        \caption[]{}
        \label{fig:oman_quint_c}
    \end{subfigure}
    \caption[ ]
    {\small \textbf{Comparison of mean spectra for continuum removed (CR) and non-CR clustering on Oman core image (offset for clarity)} (a) Cluster means on spectral data from Fig.\ \ref{fig:oman_crop_ae} (left), with CR duplicate (right) (b) Cluster means of CR data from clusters in Fig.\ \ref{fig:oman_crop_cr}. Key unique absorptions are at 1560 nm (epidote), 1750 nm (gypsum), and 1480 nm (prehnite). Important features are correlated with the vertical dotted lines.} 
    \label{fig:oman_quint}
    \vspace{-5pt}
\end{figure}

\vspace{-8pt}
\section{Experiments and Results} \label{experiments_results}
\begin{figure*}[ht!]
    \centering
    \captionsetup[subfigure]{justification=centering}
    \begin{subfigure}[t]{0.24\textwidth}
        \centering
        \includegraphics[width=\textwidth, angle=270]{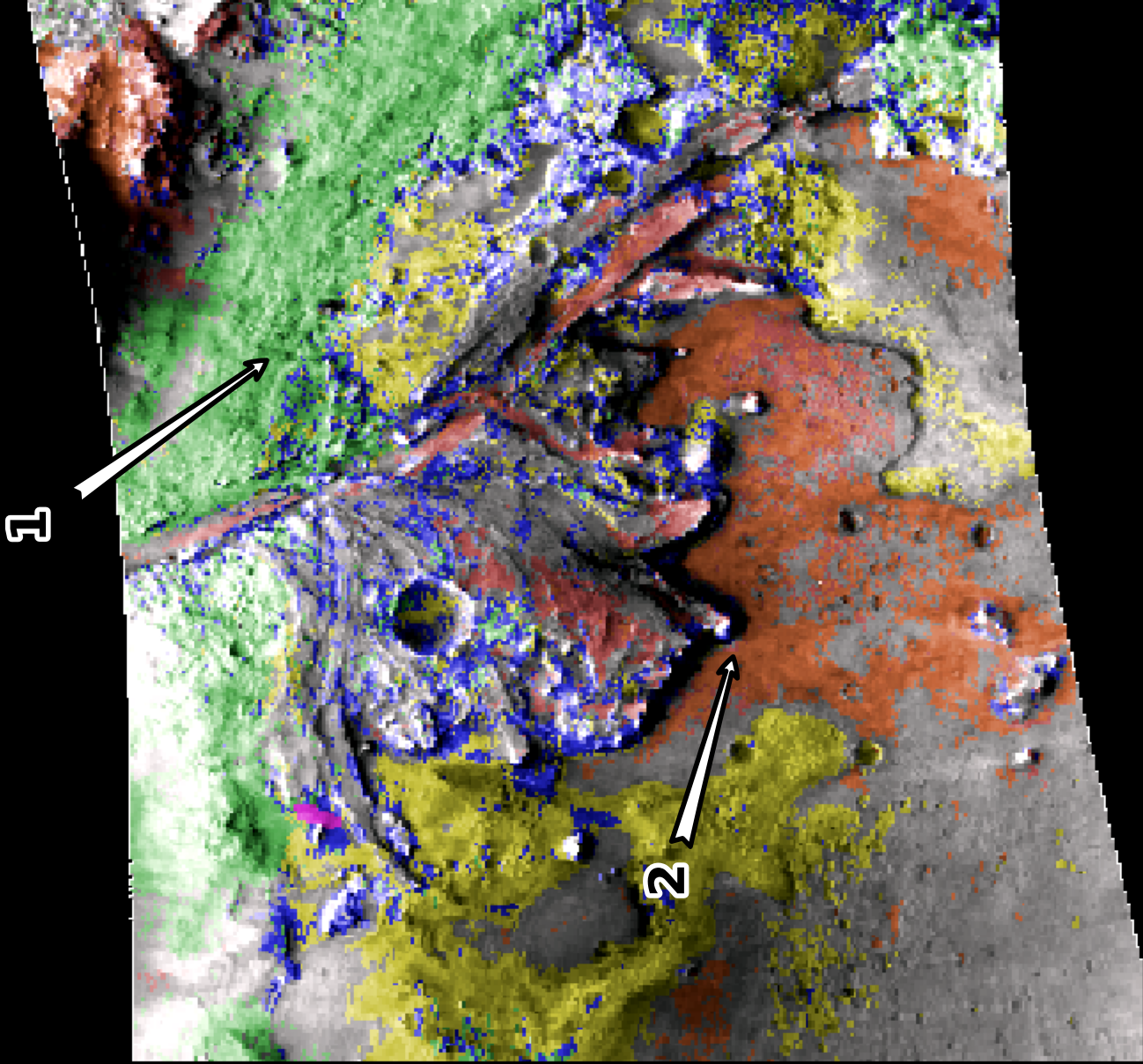}
        \caption[]
        {{\small Partial Expert Classification}}    
        \label{fig:jc_rgb}
    \end{subfigure}
    \begin{subfigure}[t]{0.24\textwidth}
        \centering
        \includegraphics[width=\textwidth, angle=270]{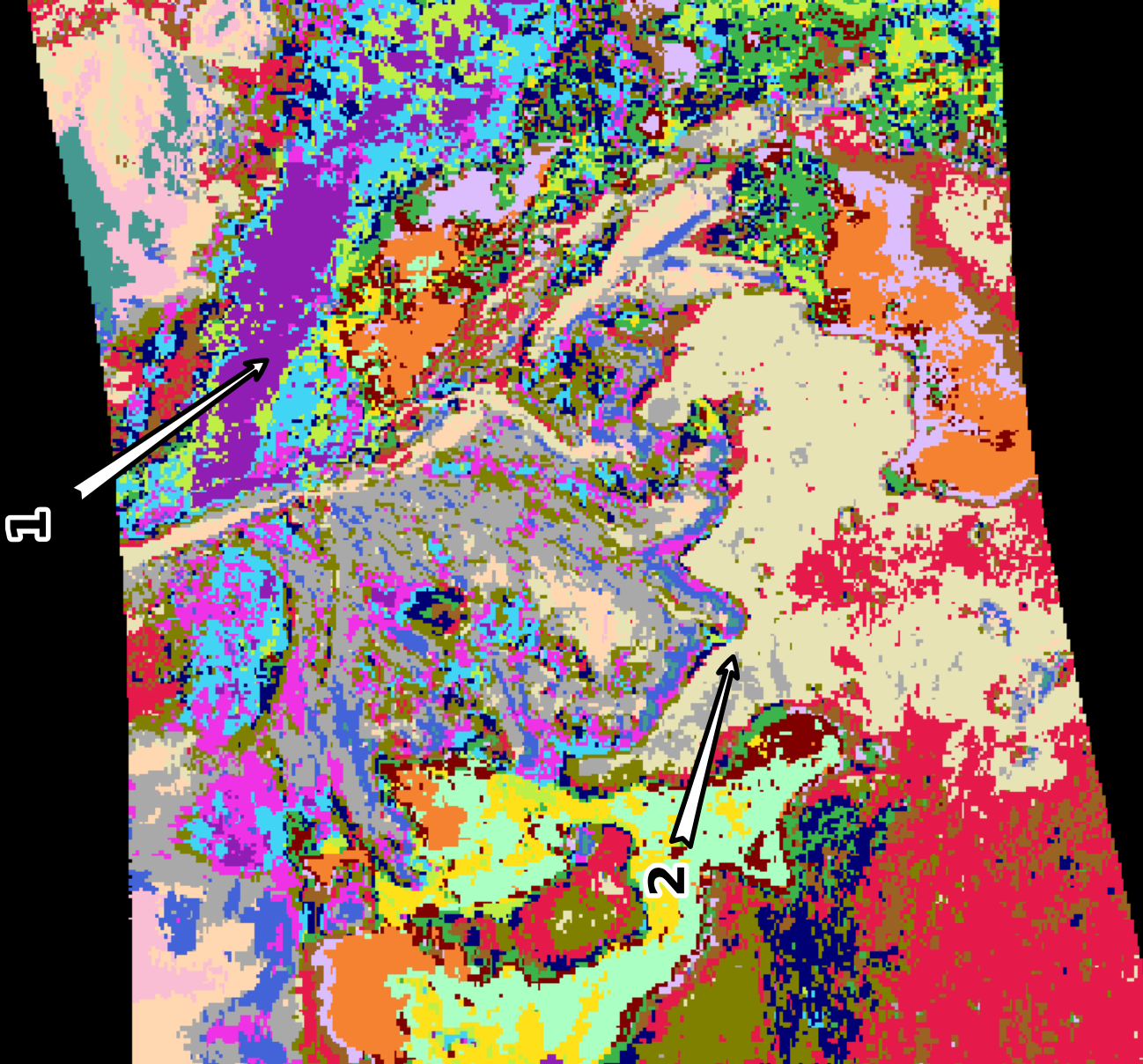}
        \caption[]
        {{\small PCA + K-Means}}    
        \label{fig:jc_pca}
    \end{subfigure}
    \begin{subfigure}[t]{0.24\textwidth}
        \centering
        \includegraphics[width=\textwidth, angle=270]{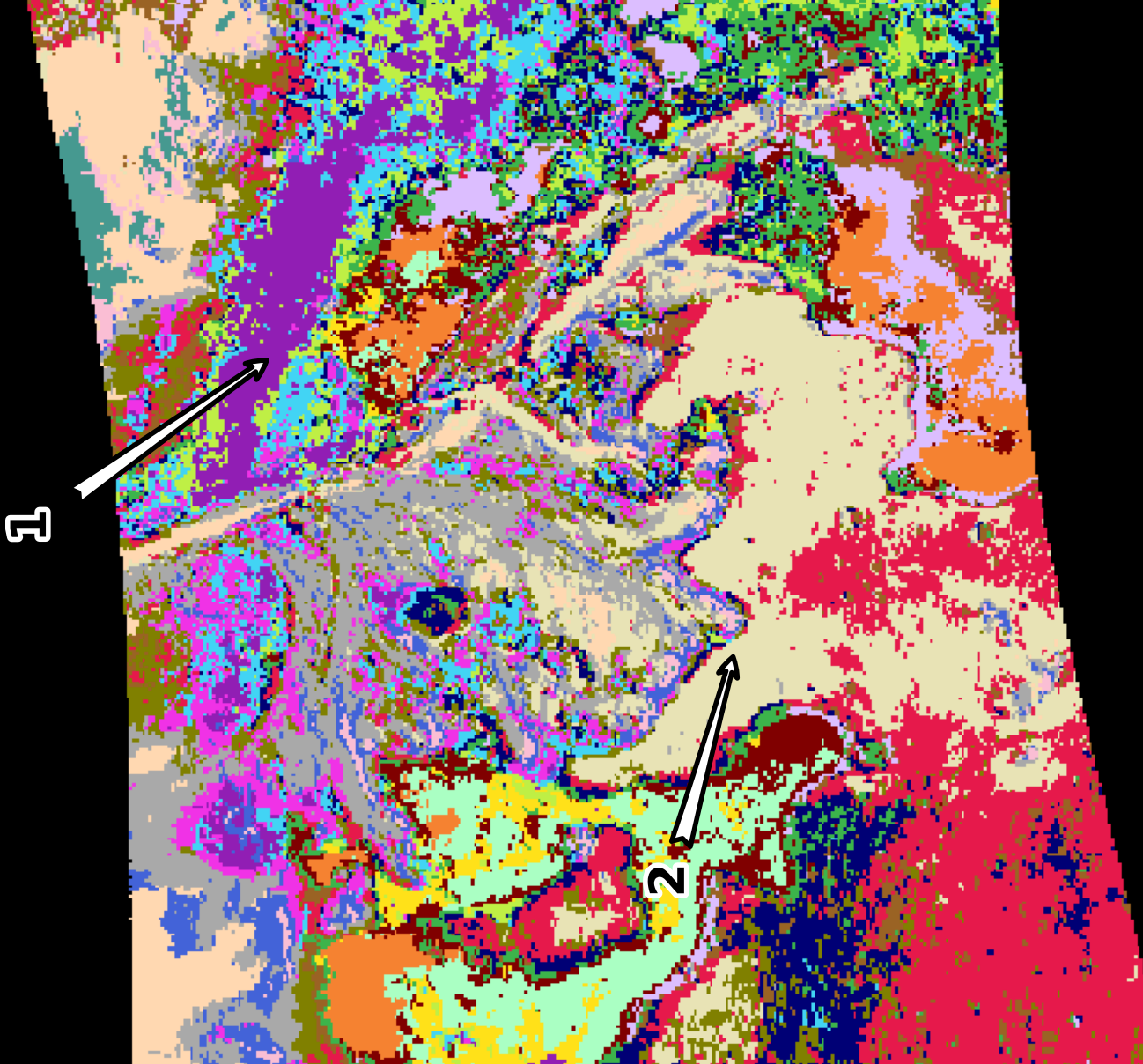}
        \caption[]
        {{\small AE + GMM on\\ Unratioed Spectral Data}}
        \label{fig:jc_ae}
    \end{subfigure}
    \begin{subfigure}[t]{0.24\textwidth}   
        \centering 
        \includegraphics[width=\textwidth, angle=270]{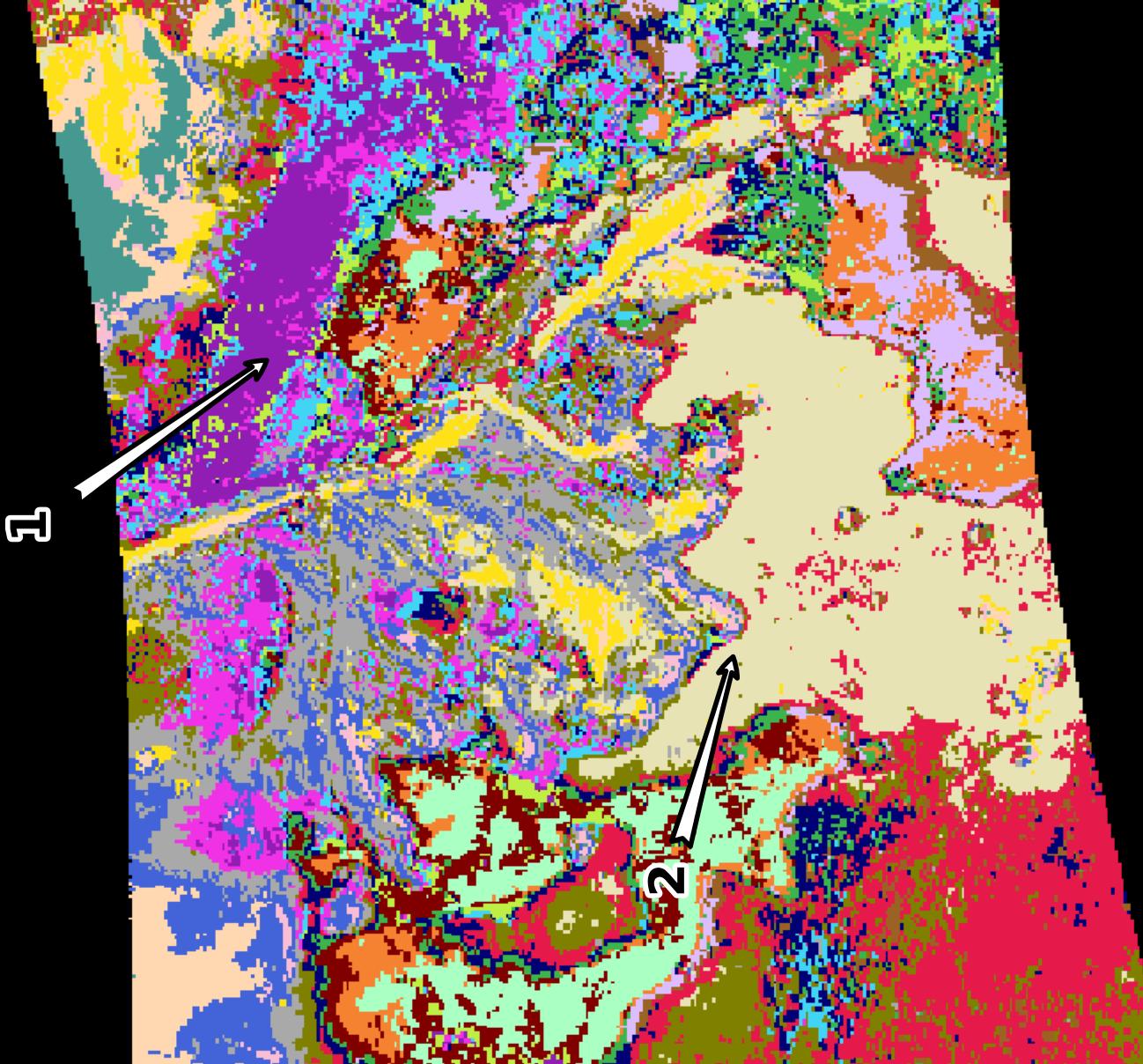}
        \caption[]
        {{\small AE + GMM on Ratioed Data}}    
        \label{fig:jc_ratio}
    \end{subfigure}
    \caption[ ]
    {\small \textbf{Map projected Jezero Crater clustering with different methods} (a) Partial expert classified mineral map overlaying greyscale image (olivine, yellow {\color[HTML]{B6B135}$\blacksquare$}; pyroxene, orange {\color[HTML]{AD5831}$\blacksquare$}; carbonate, green {\color[HTML]{51A84F}$\blacksquare$}; Fe/Mg smectite, blue {\color[HTML]{2527AD}$\blacksquare$}; silica, magenta {\color[HTML]{D22CCD}$\blacksquare$}; unclassified, gray {\color[HTML]{727070}$\blacksquare$}. 6 total classes). Note the delta feature in the top center of the image, with distinct pyroxene-bearing unit bounding its edge below (tan {\color[HTML]{E8E3B5}$\blacksquare$}, labeled 2). A spatially coherent carbonate unit branches off to the right at its top (purple {\color[HTML]{911EB4}$\blacksquare$}, labeled 1). }
    \label{fig:jc-full}
    \vspace{-10pt}
\end{figure*}

\begin{figure}
    \centering
    \begin{subfigure}[t]{0.31\textwidth}  
        \centering 
        \includegraphics[width=0.49\textwidth]{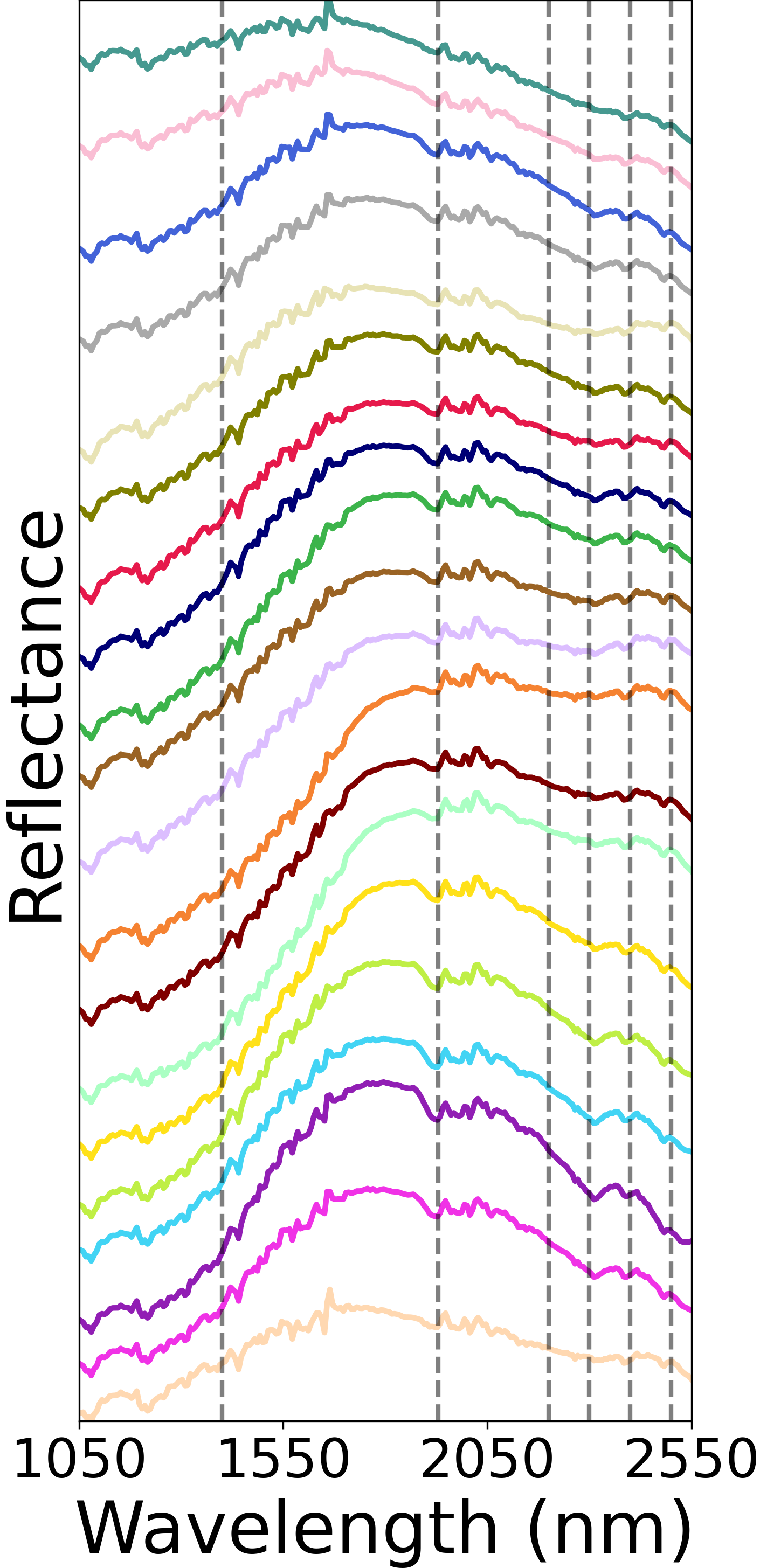}
        \includegraphics[width=0.49\textwidth]{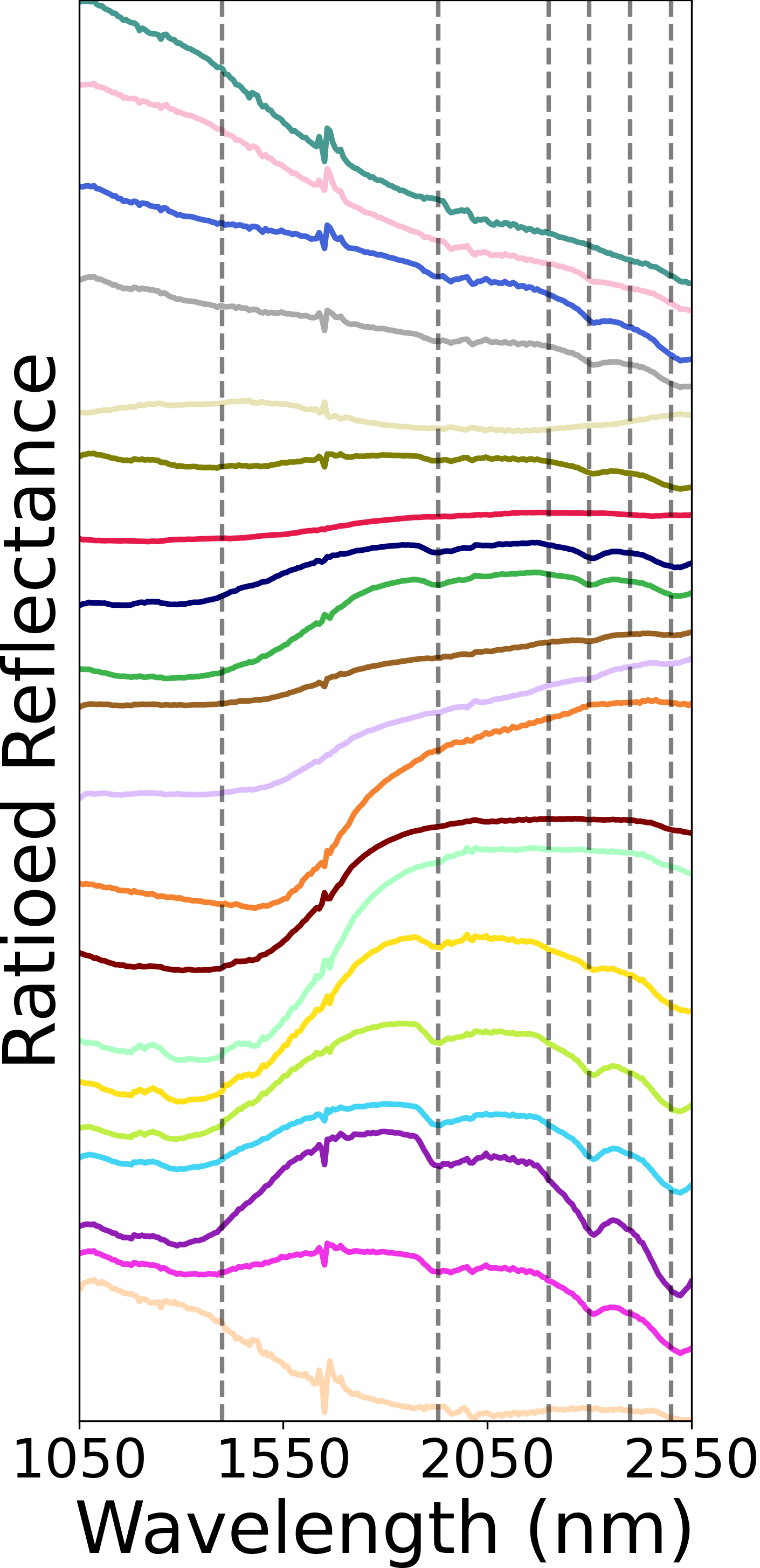}
        \caption[]{}
        \label{fig:jc_quint_b}
    \end{subfigure}
    \hfill
    \begin{subfigure}[t]{0.155\textwidth}   
        \centering 
        \includegraphics[width=0.98\textwidth]{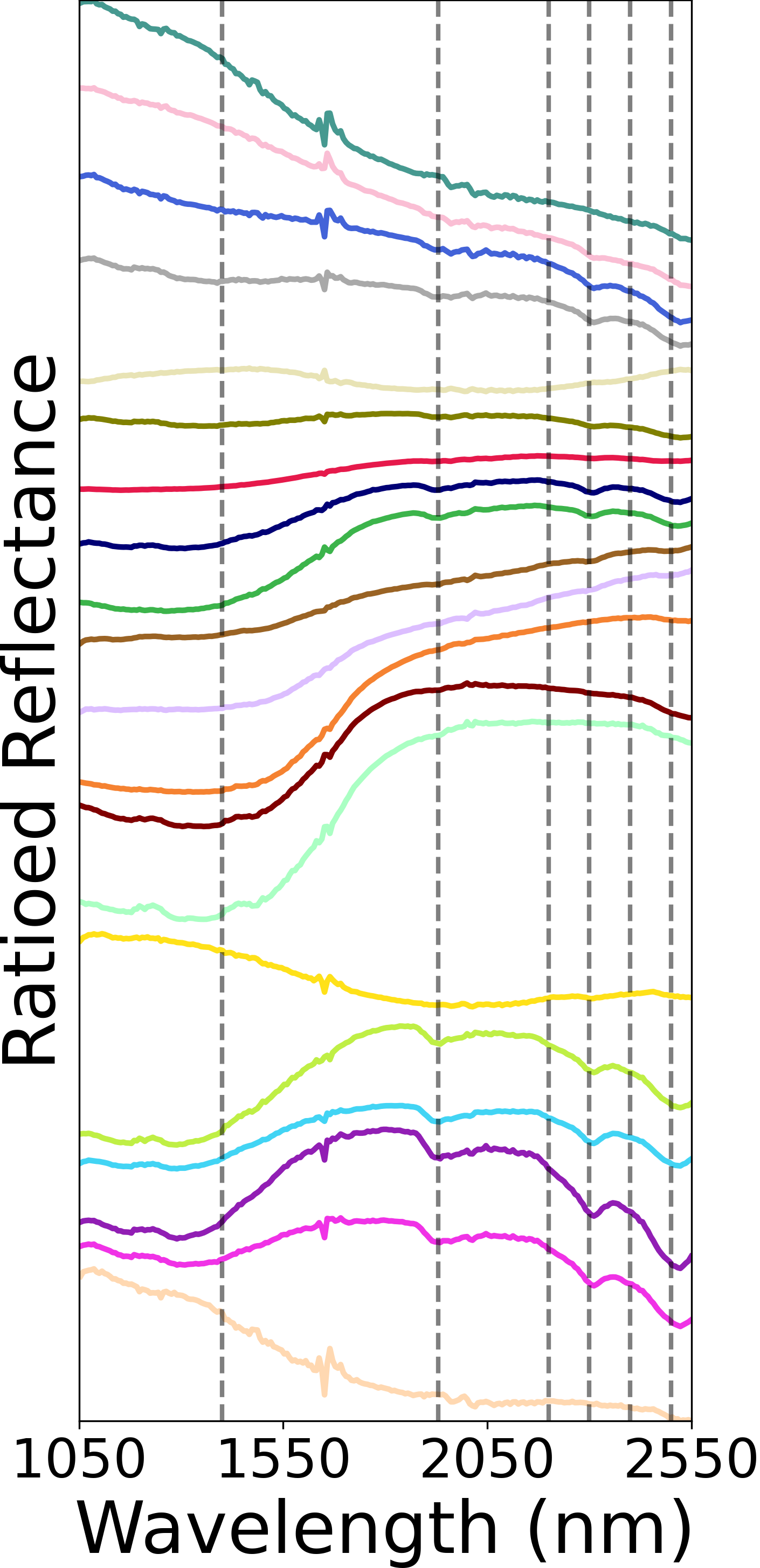}
        \caption[]{}
        \label{fig:jc_quint_c}
    \end{subfigure}
    \caption[ ]
    {\small \textbf{ Comparison of mean spectra for ratioed and unratioed clustering on Jezero Crater image (offset for clarity)} (a) Cluster means on spectral data from Fig.\ \ref{fig:jc_ae} (left) with ratioed duplicate (right) (b) Cluster means of ratioed data from clusters in Fig.\ \ref{fig:jc_ratio}. Key unique absorptions are at 1900 nm (water in minerals), 2300 nm and 2500 nm (carbonate), 2300 nm (and no 2500 nm; Fe/Mg smectite), a broad absorption from 1050 nm to 1800 nm (olivine), and a broad absorption from 1300 nm to 2300 nm (pyroxene). Important features are correlated with the vertical dotted lines.}
    \label{fig:jc_quint}
\end{figure}

We apply our methods to two datasets and show results for one image from each with additional images in the supplement. For consistency in presentation of classification images, we show the results for 20 latent embeddings and 20 clusters in Figures \ref{fig:oman_quint} and \ref{fig:jc_quint} despite better quantitative metrics for different combinations for both images (Tables \ref{tab:oman_supervised_metrics}, \ref{tab:OmanAll}, and \ref{tab:JCAll}). Colors are matched between images to maximize the pixel-wise color similarity of the largest clusters. We perform no post-processing to selectively remove redundant endmembers for classification images. The 11-class expertly labeled image for the Oman dataset (Fig.\ \ref{fig:oman_crop_gt}) and the 6-class partially classified Jezero crater image \ref{fig:jc_rgb} are displayed without color matching.

\subsection{Oman Core Evaluation}

\begin{table}[]
    \centering
    \textbf{Unsupervised Clustering Metrics: Oman}
    \vspace{2mm}
    \\
    CH Scores Using Embeddings ($\times 10^5$) 
\begin{tabular}{lccc}
\toprule
{} &   PCA &  Non-CR Data &  CR Data \\
\midrule
15 Clusters & 1.858 &          \textbf{1.016} &    0.958 \\
20 Clusters & 1.627 &          0.818 &    \textbf{0.876} \\
25 Clusters & 1.464 &          0.744 &    \textbf{0.801} \\
\bottomrule
\end{tabular}
\\
\vspace{1mm}
DB Scores Using Embeddings 
\begin{tabular}{lccc}
\toprule
{} &   PCA &  Non-CR Data &  CR Data \\
\midrule
15 Clusters & 1.239 &          \textbf{1.977} &    1.986 \\
20 Clusters & 1.254 &         \textbf{ 2.165} &    2.280 \\
25 Clusters & 1.260 &         \textbf{ 2.256} &    2.266 \\
\bottomrule
\end{tabular}
\\
\vspace{1mm}
CH Scores Using Spectral Data ($\times 10^5$) 
\begin{tabular}{lccc}
\toprule
{} &   PCA &  Non-CR Data &  CR Data \\
\midrule
15 Clusters & 0.392 &          \textbf{0.840} &    0.501 \\
20 Clusters & 0.330 &          \textbf{0.656} &    0.355 \\
25 Clusters & 0.279 &          \textbf{0.597} &    0.333 \\
\bottomrule
\end{tabular}
\\
\vspace{1mm}
DB Scores Using Spectral Data 
\begin{tabular}{lccc}
\toprule
{} &   PCA &  Non-CR Data &  CR Data \\
\midrule
15 Clusters & 5.887 &          \textbf{5.843} &    6.944 \\
20 Clusters & 6.320 &         \textbf{ 6.618} &    7.876 \\
25 Clusters & 8.284 &          \textbf{5.691} &    9.488 \\
\bottomrule
\end{tabular}
    \caption[]{\small   CH (Eq.\ \ref{eq:CH})  and DB (Eq.\ \ref{eq:DB}) scores on  embeddings $\mathbf{Z} \subseteq \mathbb{R}^{20}$ and spectral data $\mathbf{X}$ varying by number of clusters. Embeddings generated from the following methods: PCA + k-means on spectral data, AE + GMM on spectral data and CR data. The best score between non-CR data and CR data embeddings is in bold. PCA scores are included as reference. Note that PCA, Non-CR, and CR embedding spaces are all different, so scores on embeddings are not directly comparable.}
    \label{tab:OmanAll}
    \vspace{-10pt}
\end{table}

In this image, we expect to map different assemblages of the mineral groups chlorite, pyroxene, zeolite, epidote, prehnite, amphibole, and gypsum, with other minerals present but spectrally inactive in this wavelength range. Several distinct veins not clustered using PCA and k-means or mapped in the expert classification are clearly identified with our pipeline (i.e.\ multiple orange {\color[HTML]{F58231}$\blacksquare$} zeolite veins towards the bottom of the image in Fig.\ \ref{fig:oman_crop_ae}), and speckle noise abundant in the PCA case (Fig.\ \ref{fig:oman_crop_pca}) is clearly reduced. The autoencoder uniquely maps subtle cross-cutting prehnite mixing in an epidote-chlorite vein (labeled 1 in Fig.\ \ref{fig:oman_crop_gt}). The autoencoder also cleanly separates gypsum from zeolite in a large vein in the center of the image where PCA struggles (neon green {\color[HTML]{B3F348}$\blacksquare$}, slate blue {\color[HTML]{126085}$\blacksquare$} in Fig.\ \ref{fig:oman_crop_gt}, grey {\color[HTML]{A9A9A9}$\blacksquare$}, purple {\color[HTML]{911EB4}$\blacksquare$} in Fig.\ \ref{fig:oman_crop_ae}). 

The main challenge for the current implementation is separating out rare, low spatial area mineral classes. With all combinations of hyperparameters, there are no distinctive amphibole (turquoise {\color[HTML]{0BDAD5}$\blacksquare$} in expertly labeled image from Fig.\ \ref{fig:oman_crop_gt}, combination of absorptions at 1390 nm, 2320 nm, and 2390 nm) or pyroxene (pink {\color[HTML]{E807DE}$\blacksquare$} in expertly labeled image from Fig.\ \ref{fig:oman_crop_gt}, strong 1050 nm with no sharp absorptions 2200-2400 nm) clusters mapped. These classes each represent $<1\%$ of pixels in the image, and the amphibole is only present in subtle mixtures with other minerals, resulting in weak diagnostic absorptions.

Although CR exaggerates spectral features of interest, the CR-learned feature space does not seem to have stronger clustering properties. The non-CR clusters outperform the CR clusterings for the unsupervised metrics (Table \ref{tab:OmanAll}). Methods using non-CR embeddings also perform better in with the F1 supervised metrics (Table \ref{tab:oman_supervised_metrics}), though NMI and ARI are both slightly improved by CR. NMI and ARI metrics for the full implementation of GyPSUM for both spectral and CR data are substantially better than PCA + k-means results  (Table \ref{tab:oman_supervised_metrics}). Supervised metrics are comparable to results presented for a suite of unsupervised methodologies on a small subset of a Cuprite, Nevada AVIRIS image which contains fewer, arguably more distinct classes \cite{yadav_study_2019}.  



\subsection{CRISM Evaluation}

For the Jezero Crater image, we expect to map different assemblages of olivine, pyroxene, carbonate, \ce{Fe}/\ce{Mg} smectite, hydrated silica, and \ce{Al}-rich phyllosilicates. We effectively differentiate distinct units of varying olivine and pyroxene, which are primary minerals that have not been altered by interaction with water. In this scene, distinct units with variable chemistry or grain size of these spectrally active components have been identified \cite{ehlmann_identification_2009, goudge_assessing_2015, horgan_mineral_2020}. Additionally, we map varying characteristics of  carbonate/olivine mixtures (purple {\color[HTML]{911EB4}$\blacksquare$}, cyan {\color[HTML]{42D4F4}$\blacksquare$}, magenta {\color[HTML]{F032E6}$\blacksquare$} in Fig.\ \ref{fig:jc_ratio}), and various Fe/Mg smectite mixtures with both pyroxene and olivine (Fig.\ \ref{fig:jc_quint} grey {\color[HTML]{A9A9A9}$\blacksquare$}, blue {\color[HTML]{4363D8}$\blacksquare$}, dark blue {\color[HTML]{071983}$\blacksquare$}). The ratioing process better defines continuous, unaltered or very weakly altered units (i.e.\ tan {\color[HTML]{E8E3B5}$\blacksquare$} in Fig.\ \ref{fig:jc_ratio}).  

Ratioing performs better in the unsupervised clustering metrics for both embeddings and spectral data (Table \ref{tab:JCAll}). These metrics show that ratioing produces dense clusters in the learned feature space. Ratioed embeddings performed substantially better on the spectral data, and comparison with PCA and k-means shows that our methodology is less sensitive to push-broom sensor striping noise\footnote{See FRT0000634B in Supplementary Materials}. When comparing the PCA + k-means and AE + GMM on spectral data, the clustering is nearly identical while the spectral data embedding scores are consistently better (Table \ref{tab:JCAll}). This seems to indicate that the autoencoder is learning an embedding space that produces better distinct clusters, but further investigation is necessary.

With all combinations of hyperparameters we are unable to uniquely identify hydrated silica or Al-rich phyllosilicates, (which have distinct absorptions near 2200 nm) which instead appear as subtle mixtures with carbonate-dominated clusters. These minerals, and even rarer detections of jarosite and akageneite, are not abundant in the scene and have only recently been mapped exhaustively with new expertly-guided methodologies \cite{tarnas_orbital_2019, dundar_machine-learning-driven_2019}. These classes are also not mapped in the partial expert classification provided here (Fig. \ref{fig:jc_rgb}).

\begin{table}[]
    \centering
    \textbf{Unsupervised Clustering Metrics: Jezero Crater}
    \\
    \vspace{2mm}
     CH Scores Using Embeddings ($\times 10^5$) 

\begin{tabular}{lccc}
\toprule
{} &   PCA &  Spectral Data &  Ratioed  Data \\
\midrule
10 Clusters & 1.242 &          1.391 &          \textbf{1.772} \\
15 Clusters & 1.047 &          1.247 &          \textbf{1.719} \\
20 Clusters & 0.917 &          1.164 &          \textbf{1.563} \\
25 Clusters & 0.821 &          1.153 &          \textbf{1.449} \\
\bottomrule
\end{tabular}
\\
\vspace{1mm}
DB Scores  Using Embeddings 

\begin{tabular}{lccc}
\toprule
{} &   PCA &  Spectral Data &  Ratioed  Data \\
\midrule
10 Clusters & 1.098 &          1.078 &          \textbf{1.010} \\
15 Clusters & 1.168 &          1.116 &          \textbf{1.011} \\
20 Clusters & 1.178 &          1.106 &          \textbf{1.071} \\
25 Clusters & 1.248 &          \textbf{1.095} &          1.208 \\
\bottomrule
\end{tabular}
\\
\vspace{1mm}
CH Scores Using Spectral Data ($\times 10^5$) 

\begin{tabular}{lccc}
\toprule
{} &   PCA &  Spectral Data &  Ratioed  Data \\
\midrule
10 Clusters & 0.119 &          0.012 &          \textbf{0.179} \\
15 Clusters & 0.085 &          0.008 &          \textbf{0.130} \\
20 Clusters & 0.070 &          0.007 &          \textbf{0.102} \\
25 Clusters & 0.057 &          0.006 &          \textbf{0.084} \\
\bottomrule
\end{tabular}
\\
\vspace{1mm}
DB Scores Using Spectral Data 

\begin{tabular}{lccc}
\toprule
{} &   PCA &  Spectral Data &  Ratioed  Data \\
\midrule
10 Clusters & 4.545 &         15.179 &          \textbf{4.636} \\
15 Clusters & 5.657 &         15.874 &          \textbf{6.399} \\
20 Clusters & 6.727 &         16.344 &          \textbf{6.543} \\
25 Clusters & 7.126 &         15.935 &          \textbf{5.956} \\
\bottomrule

\end{tabular}
    \caption[]{ \small  CH (Eq.\ \ref{eq:CH})  and DB (Eq.\ \ref{eq:DB}) scores on embeddings $\mathbf{Z} \subseteq  \mathbb{R}^{20}$ and spectral data $\mathbf{X}$ varying by number of clusters. Embeddings generated from the following methods: PCA + k-means on spectral data, AE + GMM on spectral data and ratioed data. The best score between spectral data embeddings and ratioed data embeddings is in bold. PCA scores are included as reference. Note that PCA, unratioed, and ratioed embedding spaces are all different, so scores on embeddings are not directly comparable.}
    \label{tab:JCAll}
    \vspace{-10pt}
\end{table}

\section{Conclusions}
In this work we find that the GyPSUM pipeline effectively clusters most of the important spectral diversity in both drill-core imagery and remote-sensing imagery. Our pipeline performs comparably to other modern unsupervised classification algorithms and is relatively fast and memory efficient. Spectral ratioing of CRISM imagery unambiguously increases both clustering performance and spectral interpretability, while continuum removal results show similar clustering performance and slightly better NMI and ARI metrics. Overall, this lightweight architecture provides a relatively fast ($\sim$8.5 minutes for 4003x275x249 Oman image, $\sim$3.5 minutes for 455x751x228 Jezero Crater image), effective initial clustering for guiding in-depth work, and provides flexibility for semi-supervised learning by separating the feature extraction and clustering processes. GyPSUM enables rapid determination of distinct mineral classes across multiple imaging systems and noise profiles, demonstrating that the technique is highly generalizable. Its main shortcoming is non-identification of spectrally distinct but spatially rare ($<$ 1\%) mineral classes that can be geologically significant. The current implementation also requires user input of a spectral angle stopping condition for optional post-processing to determine a final number of clusters. Future work will include weighted sampling for clustering to improve computation time and windowing the data or hierarchical clustering to better identify spatially rare classes.

\section*{Acknowledgements}
The authors would like to thank the Oman Drilling Project and the CRISM team for access to the data used in this work. We would also like to thank Richard Murray and Sara Beery for additional comments which improved this paper. 

{\small
\bibliographystyle{ieee_fullname}
\bibliography{ms}
}

\end{document}